\documentclass[graybox]{svmult}

\usepackage{helvet}         
\usepackage{courier}        
\usepackage{type1cm}        
\usepackage{makeidx}         
\usepackage{graphicx}        
\usepackage{multicol}        
\usepackage[bottom]{footmisc}

\usepackage{amsfonts}
\usepackage{amssymb}
\usepackage{amstext}
\usepackage{amsmath,bm}

\def\be{\begin{equation}}
\def\ee{\end{equation}}
\def\ba{\begin{eqnarray}}
\def\ea{\end{eqnarray}}
\def\lb{\label}
\def\nn{\nonumber}

\def\a{\alpha}
\def\b{\beta}

\def\d{\delta}

\def\s{\sigma}

\def\bU{\overline{U}_q}

\def\1{1\!\!{\rm I}}

\def\eod{\phantom{a}\hfill \rule{2.5mm}{2.5mm}}

\def\subbbc{{\rm C}\kern-3.3pt\hbox{\vrule height4.8pt width0.4pt}\,}

\def\vac{\mid \! 0 \rangle}

\def\Ba{\bar a}
\def\Bp{\bar p}

\def\Bx{\bar x}

\begin{document}

\title*{On the 2D zero modes' algebra of the SU(n) WZNW model}
\author{Ludmil Hadjiivanov and Paolo Furlan}
\institute{
Ludmil Hadjiivanov \at 
Elementary Particle Theory Laboratory,
Institute for Nuclear Research and Nuclear Energy (INRNE),
Tsarigradsko Chaussee 72, BG-1784 Sofia, Bulgaria,
\email{lhadji@inrne.bas.bg}
\and
Paolo Furlan \at Dipartimento di Fisica, Universit\`a degli Studi di Trieste,
Strada Costiera 11, I-34014 Trieste and
Istituto Nazionale di Fisica Nucleare (INFN), Sezione di Trieste, Italy,
\email{furlan@ts.infn.it}}
\maketitle

\abstract{
A quantum group covariant extension of the chiral parts of the Wess-Zumino-Novikov-Witten (WZNW)
model on a compact Lie group $G\,$ gives rise to two matrix algebras with non-commutative entries.
These are generated by "chiral zero modes" $a^i_\a \,, \Ba^\b_j\,$ which combine, in the $2D\,$ model, into
$Q^i_j = a^i_\a \otimes \Ba^\a_j\,.$ The $Q$-operators provide important information
about the internal symmetry and the fusion ring.
Here we review earlier results about the $SU(n)\,$ WZNW $Q$-algebra and its Fock representation
for $n=2\,$ and make the first steps towards their generalization to $n\ge 3\,.$
}



\section{Introduction}
\label{intro}

The object of our study, the "zero modes", appear naturally in the splitting of the (single valued) $2D\,$ WZNW field
$G(x, \Bx) = (G^A_B (x, \bar x)) \,$ into left and right quantum group covariant chiral components
$g^A_\a (x)\,$ and ${\bar g}^\a_B (\bar x)\,.$ The latter are necessarily quasiperiodic, i.e. have monodromies:
for example, $g^A_\a (x+2\pi) = g^A_\b (x)\, M^\b_\a\,.$
The {\em chiral zero modes} $a = (a^i_\a)\,$ and $\Ba = (\Ba^\a_j)\,$ are assumed to diagonalize
the left and right monodromy matrices, respectively, so that
\be
G^A_B (x, \bar x) = g^A_\a (x)\otimes {\bar g}^\a_B (\bar x) =
u^A_i (x) \otimes {Q^i_j} \otimes {\bar u}^j_B (\bar x)\ ,\quad
Q^i_j := a^i_\a \otimes {\bar a}^\a_j
\lb{Ggg}
\ee
(summation over repeated upper and lower indices is implicitly understood),
where the chiral fields $u(x)\,$ and ${\bar u}(\Bx)\,$ have diagonal monodromies.
We call hereafter $Q:=(Q^i_j)\,$ the matrix of {\em $2D\!$ WZNW zero modes}.

The concept of chiral WZNW zero modes, classical or quantum, appeared in \cite{AF, FG, BF},
and has been developed further in \cite{HIOPT, FHIOPT, FHT6}.
The $Q$-algebra has been studied, in the $SU(2)\,$ case, in \cite{FHT2}.
It can be shown that a finite dimensional quotient of it, and the Fock representation thereof,
provide a link to the internal symmetry and the fusion of the unitary WZNW model.
We will describe below the first steps in attempt to extend this framework to $SU(n)\,,\ n\ge 3\,.$

\section{Chiral WZNW zero modes}
\label{chiral}

The (left sector) chiral zero modes' algebra ${\mathcal M}_q\,$ for the $SU(n)\,$ WZNW model at level $k\,$ has been
introduced in \cite{HIOPT}. It is generated by the $n\,$ mutually commuting operators $q^{p_j}\,$ whose product
is equal to the unit operator,
\be
q^{p_i} q^{p_j} = q^{p_j} q^{p_i}\ ,\qquad \prod_{j=1}^n q^{p_j} = 1\ ,\qquad j=1,\dots ,n\ ,
\lb{prod-p=1}
\ee
and by the entries of the $n\times n\,$ zero modes' quantum matrix $a = (a^i_\a)\,$ satisfying quadratic exchange relations,
\ba
&&a^j_\b a^i_\a\, [p_{ij}-1] = a^i_\a a^j_\b\, [p_{ij}] -\,a^i_\b\, a^j_\a \, q^{{\epsilon}_{\a\b}p_{ij}} \qquad
(\,{\rm for}\quad i\ne j \quad {\rm and}\quad\alpha\ne\beta \,)\ ,\nn\\
&&[a^j_\alpha , a^i_\alpha ] = 0\ ,\qquad a^i_\alpha a^i_\beta = q^{{\epsilon}_{\alpha\beta}}\, a^i_\beta a^i_\alpha\ ,\qquad
i, j\,, \a\,, \b = 1, \dots, n\nn\\
&&(\,\epsilon_{\a\b} = - \epsilon_{\b\a}\ ,\quad \epsilon_{\a\b} = 1\quad{\rm for}\quad \a > \b\ ,\qquad
[p] := \frac{q^p-q^{-p}}{q-q^{-1}}\,)
\lb{aa2}
\ea
the following mixed relations with $q^{p_j}\,,$
\be
\lb{ExRap}
q^{p_j}\, a_{\alpha}^i = q^{\d^i_j - \frac{1}{n}}\, a^i_\a\, q^{p_j} \qquad \Rightarrow\qquad
q^{p_{j\ell}} a_{\alpha}^i =  q^{\delta_j^i - \delta_{\ell}^i} a_{\alpha}^i \, q^{p_{j\ell}}
\quad (\, p_{j\ell} := p_j - p_\ell\, )\
\ee
and the ($n$-linear in the zero modes) inhomogeneous {\em determinant condition}
\be
\lb{Dqa=Dqp}
\frac{1}{[n]!} \, \epsilon_{i_1 \ldots i_n} \, a_{\alpha_1}^{i_1} \ldots a_{\alpha_n}^{i_n} \,
\varepsilon^{\alpha_1 \ldots \alpha_n} =: \det (a) 
= {\mathcal D}_q (p) := \prod_{i<j} [p_{ij}]\ .
\ee
The $\varepsilon$-tensor in (\ref{Dqa=Dqp}) is totally $q$-antisymmetric,
\be
\varepsilon^{\alpha_1\dots \a_i \a_{i+1}\dots \alpha_n}=
- q^{-\epsilon_{\a_i \a_{i+1}}}\,\varepsilon^{\alpha_1\dots \a_{i+1} \a_i \dots \alpha_n} \ ,
\quad i=1,\dots , n-1
\lb{q-a}
\ee
its non-zero components being given by
\be
\lb{q-eps}
\varepsilon^{\alpha_1 \ldots \alpha_n}
= q^{- \frac{n(n-1)}{4}} \, (-q)^{\ell (\sigma)}
\qquad \mbox{for} \qquad \sigma = \left( {n\ \ldots\ 1}\atop{~\alpha_1 \ldots ~\alpha_n} \right)
\in {\mathcal S}_n
\ee
(the length $\ell(\s)$ of an element $\s\,$ of the symmetric group ${\mathcal S}_n\,$ is equal to the number of
{\em inversions} which, in our notation, are the pairs $(\a_i, \a_j)$ such that $\a_i < \a_j$ for $i < j$) while
$\epsilon_{i_1 \ldots i_n} = (-1)^{\ell(\eta)}\,$ for $\eta=\left( {n\ \ldots\ 1}\atop{~i_1 \dots ~i_n} \right)\in {\mathcal S}_n\,.$

The exchange relations (\ref{aa2}) originate from
\be
{\hat R}_{12} (p) \, a_1 \, a_2 = a_1 \, a_2 \, {\hat R}_{12}\qquad\Leftrightarrow\qquad
{\hat R}^{ij}_{~i'j'} (p)\, a^{i'}_\a a^{j'}_\b = a^i_{\a'} a^j_{\b'} {\hat R}^{{\a}'{\b}'}_{~\a\b}
\lb{ExRaa1}
\ee
where ${\hat R}_{12} = P_{12} R_{12}\,,\ {\hat R}_{12} (p) = P_{12} R_{12}(p)\,,$
$P_{12}\,$ is the 
permutation matrix,
$R_{12}\,$ the Drinfeld-Jimbo quantum $R$-matrix for $U_q(s\ell(n))\,$ \cite{D, J} and
$R_{12} (p)$ the corresponding {\em dynamical} quantum $R$-matrix \cite{I2, EV2, HIOPT}.
Explicitly,
\be
q^{- \frac{1}{n}}\,{\hat R}^{\a\b}_{~\a'\b'} =  \d^\a_{\b'}\d^\b_{\a'} +
(q^{-1}- q^{-{\epsilon}_{\a\b}} )\, \d^\a_{\a'}\d^\b_{\b'}\ , \quad
\epsilon_{\a\b} =
\left\{
\begin{array}{ll}
\, {~~}1&, \quad \a > \b\\
\, {~~}0&, \quad\a = \b\\
\, - 1&,\quad \a < \b
\end{array}
\right.
\lb{hatR}
\ee
(our deformation parameter is $q=e^{-i\frac{\pi}{h}}\,$ where the {\em height} $h=k+n$) and
\ba
&&q^{- \frac{1}{n}}\,{\hat R}^{ij}_{~i'j'} (p)\, =\,
a_{ij}(p)\, \delta^i_{j'} \delta^j_{i'} + b_{ij}(p)\, \delta^i_{i'} \delta^j_{j'}\ ,\nn\\
&&a_{ii}(p) = q^{-1}\ ,\quad
a_{ij}(p) = \a (p_{ij}) \,\frac{[p_{ij}-1]}{[p_{ij}]} \ ,
\quad i\ne j\qquad (\,\a (p_{ji}) = \frac{1}{\a (p_{ij})} \,)\ ,\nn\\
&&b_{ii}(p) = 0\ ,\quad b_{ij}(p) = \frac{q^{-p_{ij}}}{[p_{ij}]}\ ,\quad i\ne j\ ,
\lb{canRp}
\lb{R}
\ea
respectively. Indeed, getting rid of the denominators in (\ref{canRp}) and using the identity
$[p-1]-q^{\pm 1} [p] = -\, q^{\pm p}\,,$ we obtain (\ref{aa2}) for $\alpha (p_{ij}) = 1\,.$

The right zero modes' algebra ${\bar{\cal M}}_q\,$ is generated by $\Ba = (\Ba^\a_i)\,$ and
$q^{\Bp_j}\,.$ The relevant relations follow from the left sector's ones according to the rules
\be
q\ \to\ q^{-1}\ ,\qquad (a^{-1})^\a_i\ \to\ \Ba^\a_i\ ,\qquad q^{p_j}\ \to \ q^{\Bp_j}
\lb{rules}
\ee
which can be justified e.g. by examining the classical chiral symplectic forms and the subsequent canonical quantization
procedure \cite{FHIOPT}. Thus $q^{\Bp_j}\,$ satisfy relations identical to (\ref{prod-p=1}) as well as mixed exchange relations
\be
q^{\bar p_j}\, \bar a_i^\a  =  q^{\d_{ij} - \frac{1}{n}}\, \bar a^\a_i\, q^{\bar p_{ij}} \qquad\Rightarrow\qquad
q^{\bar p_{j\ell}}\, \bar a^\a_i =  q^{\d_{ij} - \d_{i \ell}} \, \bar a^\a_i\, q^{\bar p_{j\ell}}\ .
\lb{bp}
\ee
The right sector counterpart of (\ref{ExRaa1}) has the form
\be
\hat R_{12}\, \bar a_1\, \bar a_2\, =\, \bar a_1 \,\bar a_2 \,{\hat {\bar R}}_{12} (\bar p)\ .
\lb{exRpR}
\ee
The fact that the {\em constant} $R$-matrices in (\ref{ExRaa1}) and (\ref{exRpR}) are the same
ensures the local commutativity of the $2D\,$ field (\ref{Ggg});
there is no such requirement however for the {\em dynamical} ones.
Inserting explicitly the $\a$-dependence in the notations of the dynamical $R$-matrices
(so that e.g. ${\hat R}_{12} (p) \equiv {\hat R}^{(1)}_{12} (p) \,$ for $\a (p_{ij}) = 1\,$ in (\ref{canRp}))
we observe that (\ref{exRpR}) becomes {\em identical} to (\ref{ExRaa1}) if we choose
$\hat{\bar R}_{12}(\bar p) =\, ^t\!{\hat R}_{12}(\bar p) \equiv {\hat R}^{(\a )}_{12} (\bar p) \,$ for
\be
\a (\Bp_{ij}) = \frac{[\Bp_{ij} + 1]}{[\Bp_{ij}-1]}\ \ (\,= \frac{1}{\a (\Bp_{ji})}\,)\ .
\lb{alf}
\ee
To this end we note that the constant $R$-matrix (\ref{hatR}) is symmetric (${\hat R}^{\a\b}_{~\a'\b'} =
{\hat R}^{\a'\b'}_{~\a\b}\,,$ i.e. ${\hat R}_{12} =\, ^t\!{\hat R}_{12}$)
and that $q^{\bar p_{ij}}\,$ commutes with ${\bar a}^\a_i {\Ba}^\b_j\,,$ cf. (\ref{bp}),
so there is no change in the argument of ${\hat {\bar R}}_{12} (\bar p)\,$ when it is moved to
the left of $\bar a_1 \bar a_2 \,$ in (\ref{exRpR}). Getting rid of the denominators, we obtain
\ba
&&\Ba_j^\b \Ba_i^\a\,[{\hat \Bp}_{ij}-1]
= \Ba_i^\a \Ba_j^\b\,[{\hat \Bp}_{ij}] - \,\Ba_i^\b\, \Ba_j^\a \,q^{{\epsilon}_{\a\b}{\hat \Bp}_{ij}} \qquad
(\,{\rm for}\quad i\ne j \quad {\rm and}\quad\alpha\ne\beta \,)\ ,\nn\\
&&[\Ba_j^\alpha , \Ba_i^\alpha ] = 0\ ,\qquad \Ba_i^\alpha \Ba_i^\beta =
q^{{\epsilon}_{\alpha\beta}}\, \Ba_i^\beta \Ba_i^\alpha\ ,\qquad \a,\b, i,j=1,\dots, n\ .
\lb{aa2barn}
\ea
That (\ref{aa2}) and (\ref{aa2barn}) coincide is a desirable result, as the left and the
right sector quantities appear in (\ref{Ggg}) on equal footing. It also suggests that the definition
of $\det (\Ba)\,$ and the condition it satisfies are identical to (\ref{Dqa=Dqp}),
up to exchanging upper and lower indices; note that (\ref{q-eps}) implies
\be
\varepsilon_{\alpha_1 \ldots \alpha_n} = \varepsilon^{\alpha_1 \ldots \alpha_n}\quad\Rightarrow\quad
\varepsilon^{\alpha_1 \ldots \alpha_n} \varepsilon_{\alpha_1 \ldots \alpha_n} = [n]!\ := [n][n-1]\dots[1]\ .
\lb{eps-norm}
\ee

The chiral matrix algebras generate Fock spaces ${\cal F}_q = {\cal M}_q\vac\,$ and ${\bar{\cal F}}_q = {\bar{\cal M}}_q\vac\,$
with vacuum vector $\vac\,$ satisfying
\be
p_{ij} \vac = (j-i) \vac = \Bp_{ij} \vac\ ,\qquad a^i_\a \vac = 0 = \Ba_i^\a \vac\quad{\rm for}\quad i\ne 1\ .
\lb{Fock}
\ee
Justification of (\ref{Fock}) can be found in \cite{FHIOPT, FH2}; we will only note here that the eigenvalues
of $p_{i i+1}\,$ and $\Bp_{i i+1}\,,\ i=1,\dots ,n\,$ play the role of {\em shifted integral $s\ell(n)$ weights}.

For $q^h = -1\,,$ the condition $[p_{ij}]\, v = 0\ $ ($i\ne j$) for some vector $v\in {\cal F}_q \otimes {\bar{\cal F}}_q\,$
implies that $p_{ij}\, v = N h\, v\,$ for some integer $N\,.$ One infers from (\ref{aa2}) (and similarly, from(\ref{aa2barn})) that
\be
[p_{ij}]\, v = 0 \ \Rightarrow\  a^i_\a a^j_\b \,v = a^j_\a a^i_\b\,v\ ,\quad
[\Bp_{ij}]\, v = 0 \ \Rightarrow\  \Ba_i^\a \Ba_j^\b \,v = \Ba_j^\a \Ba_i^\b\,v\ .
\lb{pij0}
\ee

Let e.g. ${\cal J}^{(h)}_q\,$ be the two-sided ideal of ${\cal M}_q\,$ generated by the $h$-th powers of all $a^i_\a\,$
and the $2h$-th powers of $q^{p_{ij}}\,.$ It is easy to see that the quotient
${\cal M}^{(h)}_q := {\cal M}_q / {\cal J}^{(h)}_q$ is non-trivial, due to the relation (valid for $i\ne j\,,\ \a\ne\b$)
\be
\lb{genex}
[p_{ij}-1](a^j_\b)^m a^i_\a =
a^i_\a(a^j_\b)^m [p_{ij}]- [m] (a^j_\b)^{m-1}a^i_\b \,a^j_\a\, q^{\epsilon_{\a\b} p_{ij}}
\ee
generalizing the first Eq.(\ref{aa2}) for any positive integer $m\,.$
(Eq.(\ref{genex}) is easily be proved by induction, using the $q$-number relation
$[p+m] = [p][m+1] - [p-1][m]\,.$) By a similar construction we obtain the quotient right sector zero modes' algebra
${\bar{\cal M}}^{(h)}_q\,.$ We further define restricted Fock spaces and their tensor product
\be
{\cal F}^{(h)}_q\otimes {\bar{\cal F}}^{(h)}_q  = {\cal M}^{(h)}_q \otimes {\bar{\cal M}}^{(h)}_q \vac
\lb{Fock-h}
\ee
on which the algebra of $Q$-operators will act.

\section{$Q$-algebra -- the $n=2$ case}
\label{n2}
A great simplification in the $n=2\,$ case comes from the fact that the exchange relations combine
with the determinant condition (\ref{Dqa=Dqp}), which  in this case is also bilinear in the zero modes,
to form powerful operator identities.

For $n=2\,$ and $q=e^{\pm i\frac{\pi}{h}}\,$ the chiral Fock space ${\cal F}_q\,$ carries a
representation of the $2 h^3$-dimensional {\em restricted} quantum group $\bU = {\overline U}_q(s\ell(2))\,$
generated by $E, F , K\,$ such that $E^h=0=F^h\,,\ K^{2h}=1\,$ \cite{FHT7}.
The restricted Fock space ${\cal F}^{(h)}_q\,$ is $h^2$-dimensional.
The entries of the $2D$ zero modes' matrix
\be
\lb{Q}
Q = ( Q^i_j ) = \begin{pmatrix}Q^1_1&Q^1_2\cr Q^2_1&Q^2_2\end{pmatrix} \equiv
\begin{pmatrix} A&B\cr C&D\end{pmatrix}
\ee
have the following properties \cite{FHT2}.
\begin{itemize}
\item
If $( a^i_\a )^h = 0 = ( \Ba_j^\a )^h\ \ \forall\  \a \in \{1,2\}\,,$ then $(Q^i_j)^h = 0\,.$
\item
Diagonal and off-diagonal elements of $Q\,$ commute:
\be
\lb{AB}
A B = B A \,,\ C A = A C\,,\ B D = D B\,,\ C D = D C\ .
\ee
\item
The triples $A, D, L\,$ and $B, C, N\,,$ generate two commuting $\bU\,$ algebras:
\ba
&&[A , D] = [ L] \ ,\ \ L A = q^2 A L\ ,\ \  L D = q^{-2} D L\ ,\ \
L^{\pm 1} := - \, q^{\pm p} \otimes q^{\pm{\bar p}}\nn\\
&&[B , C ] = [ N]\ , \ \  N B = q^2 B N\ ,\ \ N C = q^{-2} C N\ , \ \
N^{\pm 1} :=  - \, q^{\pm p} \otimes q^{\mp {\bar p}}\nn\\
&&A^h = D^h = 0 = B^h = C^h\ ,\quad L^{2h} = 1 = N^{2h}\quad (p=p_{12}\,,\ \Bp=\Bp_{12})\ .\quad\quad
\lb{triples}
\ea
\item
The vacuum representation of the off-diagonal $Q$-algebra is one-dimensional:
\be
B \vac = 0 = C \vac \ ,\quad N \vac = - \vac \qquad (\,\Rightarrow\ [N]\vac = 0\,)\ .
\lb{BCvac}
\ee
\item
The diagonal $Q$-algebra generates an indecomposable representation of $\bU\,$
(a {\em Verma module} \cite{FGST1, FGST2} ${\cal V}_1^+ , \dim {\cal V}_1^+ = h\, ,$ with a $1$-dimensional submodule),
\ba
&&A \mid m \rangle = [m+1] \mid m + 1 \rangle\ ,\quad
D \mid m \rangle = [m+1] \mid m -1\rangle\quad (\,D \vac = 0\,)\ ,\nn\\
&&(L + \,q^{2(m+1)}) \mid m \rangle = 0\quad
{\rm for}\ \ \mid m \rangle := \frac{A^{m}}{[m]!} \vac \ ,\quad m=0,\dots , h-1\ .\qquad
\lb{m}
\ea
\item
The invariant hermitean scalar product on (\ref{m}) 
(s.t. $A^\dagger = D\,,\ L^\dagger = L^{-1}$) is semidefinite,
the $\bU$-invariant subspace ${\mathbb C}\!\mid\! h-1 \rangle\subset{\cal V}_1^+ \,$ being isotropic:
\be
( m' \mid m ) = [m+1]\,\d_{m m'} \equiv \frac{\sin{(m+1)\frac{\pi}{h}}}{\sin{\frac{\pi}{h}}}\, \d_{m m'}\ ,
\quad m=0,\dots , h-1\ .
\lb{scpr}
\ee
\end{itemize}
Note that the dimension of the quotient space ${\cal V}_1^+ / \{ {\mathbb C}\!\mid\! h-1 \rangle \}\,$
coincides with the number $h-1 = k+1\,$ of (integrable) sectors in the unitary $\widehat{su}(2)_k\,$ WZNW model \cite{DFMS}.
This is a manifestation of a much deeper result providing an interpretation analogous to covariant quantization of
gauge theories \cite{FHT2}. Without going into details, we would like to call special attention to the fact that only the
{\em diagonal} entries of the matrix $Q\,$ (\ref{Q}) are represented non-trivially. It guarantees, together with (\ref{ExRap})
and (\ref{bp}), that the eigenvalues of $p\,$ and $\Bp\,$ on the diagonal $Q$-vectors $\mid m \rangle\,$ (\ref{m}) coincide.

\section{$Q$-algebra -- the general $n$ case}
\label{genn}

The general $n\,$ case is much harder to explore, partly because the $n$-linear determinant conditions
for the chiral zero modes should be considered for $n\ge 3\,$ separately from the quadratic exchange relations.
For this reason we will only comment below the extensions to higher $n\,$ of the first two points
listed in Section \ref{n2} for $n=2\,,$ leaving the rest for a future work.

It turns out that the generalization of the first one is straightforward.

\smallskip

\noindent
{\bf Proposition~} If $\ ( a^i_\a )^h = 0 = ( \Ba_j^\a )^h\quad \forall\,\a \in \{ 1, \dots, n \}\,,\,$
then $\ (Q^i_j)^h = 0\,.$

\smallskip

\noindent
{\bf Proof~} The indices $i\,$ and $j\,$ play no role here; introducing the "$\a$-components"
$Q_\a := a^i_\a\,\otimes\, \Ba^\a_j\,$ (no summation in $\a\,$ is assumed) of $Q^i_j = \sum_{\a = 1}^n Q_\a\,,$ we have
\be
(Q_\a)^h = ( a^i_\a )^h \otimes ( \Ba_j^\a )^h = 0\ ,\qquad
Q_\a\, Q_\b = a^i_\a a^i_\b \otimes \Ba_j^\a \Ba_j^\b = q^{2\epsilon_{\a\b}}\, Q_\b\, Q_\a\ .\quad
\lb{Qhn}
\ee
We will perform the proof by induction in $n\,,$ observing that
\be
Q_\a \,(Q_1 + \dots + Q_{\a-1}) = q^2 (Q_1 + \dots + Q_{\a-1})\, Q_\a\ ,\quad \a = 2,\dots ,n\ .
\lb{Qrh1}
\ee
The calculation is based on the $q$-binomial identity (in fact, the case $n=2$)
\be
Q_2 Q_1 = q^2 Q_1 Q_2\quad\Rightarrow\quad (Q_1+Q_2)^m = \sum_{r=0}^m \left({m\atop r}\right)_+ Q_1^r Q_2^{m-r}
\lb{qbin}
\ee
where
$\left({m\atop r}\right)_+ = \frac{(m)_+!}{(r)_+! (m-r)_+!}\,,\ (r)_+! = (r)_+ \dots(1)_+\,,\ (r)_+ = \frac{q^{2r}-1}{q^2-1}\,,$
implying
\be
(Q_1 + Q_2)^h = (Q_1)^h + \sum_{r=1}^{h-1} \left({h\atop r}\right)_+ Q_1^r Q_2^{h-r} + (Q_2)^h = 0
\lb{Ah}
\ee
(Eq. (\ref{qbin}) can be proved by induction in $m$). Eqs. (\ref{Qrh1}), (\ref{qbin}) and (\ref{Ah}) imply
\be
(Q_1 + \dots + Q_\a )^h = (Q_1 + \dots + Q_{\a-1} )^h + (Q_\a)^h = (Q_1 + \dots + Q_{\a-1} )^h\ ,
\lb{Qrh}
\ee
etc. The following general formula can be proved by induction as well:
\be
\left( \sum_{\a = 1}^n Q_\a \right)^h = \sum_{\a = 1}^n (Q_\a)^h +\
(h)_+! \!\!\!\!\!\!\!\!\!\!\!\!\! \sum_{{m_1+m_2+\dots +m_n = h}\atop{0\le m_i \le h-1}}
\frac{(Q_1)^{m_1}}{(m_1)_+!} \frac{(Q_2)^{m_2}}{(m_2)_+!} \dots \frac{(Q_n)^{m_n}}{(m_n)_+!} = 0\ .
 \lb{Paolo}\eod
\ee

\smallskip

In compliance with the final remark of Section \ref{n2}, we will make the following

\smallskip

\noindent
{\bf Conjecture ~} {\em Any $Q$-monomial containing off-diagonal entries of $\ Q\,$ annihilates the vacuum vector.}

\smallskip

Recall that in the $n=2\,$ case this property is valid, due to the general fact (following from (\ref{Fock}))
that $Q^i_j\! \vac = 0\,$ for $i\ne j\,$ and the commutativity of the diagonal and off-diagonal entries of $Q\,$ (\ref{AB})
which however doesn't hold in general but is replaced by the following corollaries of (\ref{aa2}) and (\ref{aa2barn}).

\smallskip

\noindent
{\bf Lemma 1~} {\em The entries of $Q\,$ belonging to the same row or column commute:}
\be
[Q^j_i , Q^\ell_i ] = 0 = [Q^i_j , Q^i_\ell ]\ .
\lb{QQcomm}
\ee

\smallskip

\noindent
{\bf Proof~} It is sufficient to explore the case in (\ref{QQcomm}) when the different indices ($j\,$ and $\ell$)
are carried by the left sector variables since the bar quantities satisfy identical relations.
We obtain (assuming implicitly that equal upper and lower {\em greek\,} i.e. quantum group, indices are summed over
all admissible values from $1\,$ to $n\,,$ if no restrictions are indicated under a summation symbol)
\ba
&&[p_{\ell j}-1] \, Q^j_i\, Q^\ell_i = [p_{\ell j}-1] \,(a^j_\b\otimes \bar a^\b_i) (a^\ell_\a\otimes \bar a^\a_i) =
[p_{\ell j}-1] \,a^j_\b a^\ell_\a \otimes \bar a^\b_i \bar a^\a_i = \nn\\
&&=[p_{\ell j}-1] \,\sum_\a a^j_\a a^\ell_\a \otimes \bar a^\a_i \bar a^\a_i
+ \sum_{\a\ne\b} [p_{\ell j}-1] \,a^j_\b\, a^\ell_\a \otimes \bar a^\b_i \bar a^\a_i =\lb{QjiQii1}\\
&&= [p_{\ell j}-1] \,\sum_\a a^\ell_\a a^j_\a \otimes \bar a^\a_i \bar a^\a_i +
\sum_{\a\ne\b} \left( a^\ell_\a a^j_\b\,[p_{\ell j}] - \,a^\ell_\b\, a^j_\a \,q^{{\epsilon}_{\a\b}p_{\ell j}}\right) \otimes \bar a^\b_i \bar a^\a_i =\nn\\
&&= [p_{\ell j}-1] \,\sum_\a a^\ell_\a a^j_\a \otimes \bar a^\a_i \bar a^\a_i +
\sum_{\a\ne\b} a^\ell_\b\, a^j_\a \, \left(q^{{\epsilon}_{\a\b}} {[p_{\ell j}]} - q^{\epsilon_{\a\b} p_{\ell j} } \right)
\otimes \bar a^\b_i \bar a^\a_i =\nn\\
&&= [p_{\ell j}-1] \, a^\ell_\b\, a^j_\a \otimes \bar a^\b_i \bar a^\a_i = [p_{\ell j}-1] \,Q^\ell_i\, Q^j_i
\quad{\rm i.e.,}\quad [p_{\ell j}-1] \,[ Q^j_i\,, Q^\ell_i ] = 0\qquad
\nn
\ea
(we have applied (\ref{aa2}), exchanged the dummy indices $\a\,$ and $\b\,$ in a term on the fourth line and then used the identity
$q^\epsilon [ p ] - q^{\epsilon p}= [p-1]\,$ for $\epsilon = \pm 1$). The first relation (\ref{QQcomm}) $[Q^j_i , Q^\ell_i ] = 0\,$ follows
since, by exchanging the upper (left sector) indices $j\,$ and $\ell\,,$ we can also derive that
\be
[p_{j\ell}-1]\,[Q^\ell_i\,, Q^j_i ] = [p_{\ell j}+1]\,[Q^j_i\,, Q^\ell_i ] = 0\ ,
\lb{ij-exch}
\ee
and there is no vector on which the operators $[p_{\ell j} + 1]\,$ and $[p_{\ell j} - 1]\,$ vanish simultaneously.
In a similar way one obtains from (\ref{aa2barn}) that $[Q_j^i \,, Q_\ell^i] = 0\,.$\eod

\smallskip

\noindent
{\bf Lemma 2~} {\em The entries of $Q\,$ belonging to different rows and columns satisfy}
\ba
&&( [p_{ij}-1]\otimes [\Bp_{\ell m} ]- [p_{ij} ] \otimes [\Bp_{\ell m}-1])\, Q^i_\ell \, Q^j_m\quad
(\, \equiv [p_{ij}-\Bp_{\ell m} ]\, Q^i_\ell \, Q^j_m \,)  =\nn\\
&&= [p_{ij}-1]\otimes [\Bp_{\ell m} ]\,Q^j_\ell\, Q^i_m - [p_{ij} ]\otimes [\Bp_{\ell m} -1 ]\,  Q^i_m\, Q^j_\ell
\quad (\,i\ne j\,,\ \ell\ne m\,) \ .\qquad
\lb{QQijlm}
\ea

\smallskip

\noindent
{\bf Remark ~} Below we will make use of the following $q$-identities:
\ba
&&[p \pm 1] \otimes [\Bp\, ] - [p\, ] \otimes [\Bp \pm 1] = \mp \, [ p-\Bp\, ]
:= \mp \, \frac{q^p\otimes q^{-\Bp} - q^{-p}\otimes q^{\Bp}}{q-q^{-1}}\ ,\nn\\
&&[p \pm 1] \otimes [\Bp\, ] - [p\, ] \otimes [\Bp \mp 1] = \pm \, [ p+\Bp\, ]
:= \pm \, \frac{q^p\otimes q^{\Bp} - q^{-p}\otimes q^{-\Bp}}{q-q^{-1}}\ ,\nn\\
&&[p\,] \otimes q^{\epsilon \Bp} - q^{\epsilon p}\otimes [ \Bp\, ] =: [ p - \Bp\,]\ ,\quad \epsilon = \pm 1\ .
\lb{ids}
\ea

\smallskip

\noindent
{\bf Proof~}
Eq. (\ref{QQijlm}) is suggested by (\ref{ExRaa1}) and (\ref{exRpR}), (\ref{alf}) implying
\be
{\hat R}^{ij}_{~i'j'} (p)\,Q^{i'}_\ell\,Q^{j'}_m = Q^i_{\ell'}\,Q^j_{m'}\,({\hat R}^{(\a)})^{\ell' m'}_{~\ell m}(\bar p)
\lb{RQQ}
\ee
but can be also verified directly with the help of (\ref{aa2}), (\ref{aa2barn}) and (\ref{ids}):
\ba
&&[p_{ij}-1]\otimes [\Bp_{\ell m} ]\,Q^j_\ell\, Q^i_m - [p_{ij} ]\otimes [\Bp_{\ell m} -1 ]\,  Q^i_m\, Q^j_\ell = \nn\\
&&=[p_{ij}-1]\otimes [\Bp_{\ell m} ]\,\sum_\a a^j_\a\,a^i_\a \otimes \Ba^\a_\ell\,\Ba^\a_m +
\sum_{\a\ne\b} ([p_{ij}]\,a^i_\a\,a^j_\b - q^{\epsilon_{\a\b} p_{ij}} a^i_\b \, a^j_\a ) \otimes
[\Bp_{\ell m}]\, \Ba^\b_\ell\,\Ba^\a_m -\nn\\
&&- \, [p_{ij} ]\otimes [\Bp_{\ell m} -1 ]\,\sum_\a a^i_\a\,a^j_\a \otimes \Ba^\a_m\,\Ba^\a_\ell -
\sum_{\a\ne\b} [p_{ij}]\,a^i_\b\,a^j_\a \otimes
(\, [ \Bp_{\ell m} ]\, \Ba^\a_\ell\,\Ba^\b_m - q^{\epsilon_{\a\b} \Bp_{\ell m}} \Ba^\b_\ell\,\Ba^\a_m \, ) = \nn\\
&&= [p_{ij}-\Bp_{\ell m} ]\,\sum_\a a^i_\a\,a^j_\a \otimes \Ba^\a_\ell\,\Ba^\a_m
+ \sum_{\a\ne\b}\, (\, [p_{ij}]\otimes q^{\epsilon_{\a\b}\Bp_{\ell m}} -
q^{\epsilon_{\a\b} p_{ij}} \otimes [ \Bp_{\ell m} ]\, )\, a^i_\b\,a^j_\a \otimes \Ba^\b_\ell\,\Ba^\a_m  = \nn\\
&&= [p_{ij}-\Bp_{\ell m} ]\, Q^i_\ell \, Q^j_m \qquad (\, i\ne j\,,\ \ell\ne m \,)\ .
\lb{QQijlm1} \eod
\ea
Let us see what the above two Lemmas tell us in the cases involving diagonal entries of $Q\,.$
Eq. (\ref{QQcomm}) implies that
\be
[Q^j_i , Q^i_i ] = 0 = [Q^i_j , Q^i_i ]\ ,
\lb{same}
\ee
while Eq. (\ref{QQijlm}) gives rise to the following relations valid for $i\ne j\ne \ell \ne i\,$
(which is only possible if $n\ge 3$):
\ba
&&[p_{ij}-1]\otimes [\Bp_{i\ell} ]\,Q^j_\ell\, Q^i_i = [p_{ij} ]\otimes [\Bp_{i\ell} +1 ]\,  Q^i_i\, Q^j_\ell -
[p_{ij}+\Bp_{i\ell} ]\, Q^i_\ell \, Q^j_i\ ,\nn\\
&&[p_{ij}]\otimes [\Bp_{i\ell}-1 ]\,Q^j_\ell\, Q^i_i = [p_{ij} +1 ]\otimes [\Bp_{i\ell} ]\,  Q^i_i\, Q^j_\ell -
[p_{ij}+\Bp_{i\ell} ]\, Q^j_i \, Q^i_\ell\ .
\lb{no1}
\ea
So an off-diagonal $Q$-operator can jump over a diagonal one, except in cases when the $p$-dependent
coefficients in the left-hand sides of the two identities (\ref{no1}) vanish simultaneously
(note that the last terms of (\ref{no1}) only contain off-diagonal $Q$-operators).
Moreover, if $[p_{ij}]\, v = 0\,$ or $\,[\Bp_{i\ell}]\, v = 0\,,$ then
\be
[p_{ij}]\, v = 0\ \Rightarrow\ Q^j_\ell\, Q^i_i\, v = Q^i_\ell\, Q^j_i \, v\ ,\quad
[\Bp_{i\ell}]\, v = 0\ \Rightarrow\  Q^j_\ell\, Q^i_i \, v = Q^j_i\, Q^i_\ell\, v
\lb{pij-or-pil}
\ee
by (\ref{pij0}), so the only obstacle arises when we apply (\ref{no1}) to vectors $v\,$ satisfying
\be
[p_{ij}-1] \,v = 0 = [\Bp_{i\ell}-1 ]\,v\qquad{\rm for}\qquad i\ne j\ne \ell \ne i\ .
\lb{probl1}
\ee
The above facts open the possibility to prove the Conjecture by induction in the number of diagonal $Q$-operators
applied to the vacuum, starting with $v_0 = \vac\,$ and $v_1 = Q^1_1\vac\,.$ To find out if and when (\ref{probl1})
occurs, we need to explore the space of diagonal $Q$-vectors
${\cal F}^{diag} = \{ v \ |\  v = P(Q^n_n\,,\dots , Q^1_1) \vac \}\,$
and its subspace ${\cal F}'\subset {\cal F}^{diag}\,$ that is annihilated by all off-diagonal elements,
$Q^r_s\, {\cal F}' = 0\,,\ r\ne s\,.$ (In these terms our conjecture is equivalent to
${\cal F}' \stackrel{(?)}{=} {\cal F}^{diag}$.) The exchange relations for diagonal elements
that follow from (\ref{QQijlm})
\ba
&&[p_{ij} ]\otimes [\Bp_{ij} +1 ]\,  Q^i_i\, Q^j_j - [p_{ij}-1]\otimes [\Bp_{ij} ]\,Q^j_j\, Q^i_i =
[p_{ij}+\Bp_{ij} ]\, Q^i_j \, Q^j_i\ ,\lb{no2}\\
&&[p_{ij} +1 ]\otimes [\Bp_{ij} ]\,  Q^i_i\, Q^j_j - [p_{ij}]\otimes [\Bp_{ij}-1 ]\,Q^j_j\, Q^i_i =
[p_{ij}+\Bp_{ij} ]\, Q^j_i \, Q^i_j\quad (\,i\ne j\,)\ .
\nn
\ea
imply (as the eigenvalues of $p_{ij}\,$ and $\Bp_{ij}\,$ on ${\cal F}^{diag}\,$ are equal)
\be
[p_{ij}+1]\,Q^i_i \, Q^j_j \,\approx\, [p_{ij}-1]\, Q^j_j\, Q^i_i
\lb{QQ-diag}
\ee
where the "weak equality" sign refers to an identity that holds on ${\cal F}'\,.$

As already mentioned, these are just the first steps in our study of the $Q$-algebra and its vacuum representation
for $n\ge 3\,.$ The obvious immediate tasks are the completion of the proof of the diagonality conjecture
and the description of ${\cal F}^{diag}\,.$ To this end, one should take next into account (besides the bilinear
exchange relations) the $n$-linear determinant condition
(which suggests a basis in ${\cal F}^{diag}\,$ labelled by $su(n)\,$ Young diagrams \cite{Ful})
and also some trilinear relations following from the chiral structure of the $Q$-operators. Together
with $(Q^i_i)^h = 0\,$ (\ref{Paolo}) and (\ref{QQ-diag}), the latter seem to imply the finite dimensionality of
${\cal F}^{diag}\,.$

\section{Discussion and outlook}
\label{discussion}

It would be intriguing to look for a possible connection of the diagonal $Q$-algebra with the algebra of the
(phase model) {\em "hopping operators"} $\{ Q_1, \dots , Q_n \}$
on a circle (also called {\em "affine local plactic algebra"}). The latter is characterized by the relations
\ba
&&[Q_i , Q_j] = 0 \ ,\quad {\rm if}\ \ i\ne j\pm 1\ {\rm mod}\, n\nn\\
&&Q_i\, Q^2_j = Q_j\, Q_i\, Q_j \ ,\quad Q_i^2\, Q_j = Q_i\, Q_j\, Q_i \ ,\quad  {\rm if}\ \
i = j + 1\ {\rm mod}\, n\qquad
\lb{hop}
\ea
and provides a description of the (unitary) $\widehat{su}(n)_k\ $ affine fusion ring \cite{KS, Wa2012}.
In contrast to our (diagonal) $Q$-algebra, it {\em does not} depend explicitly on the level $k\,$
which only labels its representations. Although it is clear from the outset that the two algebras are not isomorphic,
relations (\ref{hop}) can suggest the correct procedure needed to obtain the physical subquotient space for general $n\,.$

\begin{acknowledgement}

The authors have benefited from discussions with I. Todorov and T. Popov.
They thank the organizers of the 10th International Workshop
"Lie Theory and Its Applications in Physics" (LT-10, 17-23 June 2013) held in Varna, Bulgaria for the excellent conditions and
the stimulating and friendly atmosphere.
The work of L.H. has been supported in part by INFN, Sezione di Trieste, Italy and of P.F., by
the Italian Ministry of University and Research (MIUR).


\end{acknowledgement}

\end{document}